\documentclass[10pt,preprint]{aastex}
\usepackage{amsmath,amssymb}

\def\bmath#1{{\bf #1}}

\def\pseudofigureone#1#2#3#4{
\begin{figure}
\plotone{#3}
\caption{#4
\label{#1}}
\end{figure}}

\usepackage[numberedappendix]{emulateapj5}
\usepackage{apjfonts}

\journalinfo{The Astrophysical Journal, 625: L115--L118, 2005 June 1; 
\rm preprint NORDITA-2004-106; 
e-print~{\tt astro-ph/0412594}}

\def\pseudofigureone#1#2#3#4{{
\refstepcounter{figure}
\label{#1}
\vskip2mm
\plotone{#2}
\vskip2mm
\footnotesize\def\baselinestretch{1.0}
\begin{minipage}{\columnwidth}
{\scshape ~~Fig.}\space\thefigure.--- #4
\end{minipage}
\vskip2mm
}}

\def\exref#1{(\ref{#1})}

\def\eqref#1{Eq.~(\ref{#1})}

\def\figref#1{Fig.~\ref{#1}}

\def\const{{\rm const}}

\def\bea{\begin{eqnarray}}
\def\eea{\end{eqnarray}}

\def\({\left(}
\def\){\right)}
\def\<{\langle}
\def\>{\rangle}
\def\lt{\left}
\def\rt{\right}
\def\bl{\bigl}
\def\br{\bigr}

\def\dd{\partial}

\def\vdel{\bmath{\nabla}}

\def\vx{\bmath{x}}

\def\vu{\bmath{u}}

\def\vB{\bmath{B}}

\def\vf{\bmath{f}}
\def\vF{\bmath{F}_{\rm visc}}
\def\vS{\bmath{\hat S}}

\def\cs{c_{\rm s}}
\def\nueff{\nu_{\rm eff}}

\def\ld{l_{\nu}}
\def\lf{l_0}

\def\lB{l_B}
\def\kf{k_0}

\def\usq{{\langle u^2 \rangle}}
\def\dusq{{\langle|\vdel\vu|^2\rangle}}

\def\Re{{\rm Re}}
\def\Rm{{\rm Rm}}
\def\Rmc{{\rm Rm}_{\rm c}}
\def\Rel{{\rm Re}_\lambda}
\def\Pm{{\rm Pm}}
\def\Pmc{{\rm Pm}_{\rm c}}

\slugcomment{\today}
\shorttitle{ONSET OF SMALL-SCALE DYNAMO} 
\shortauthors{SCHEKOCHIHIN ET AL.}

\begin{document}

\title{The Onset of a Small-Scale Turbulent Dynamo at Low Magnetic Prandtl Numbers}
\author{A.~A.~Schekochihin,\altaffilmark{1} 
N.~E.~L.~Haugen,\altaffilmark{1,2}
A.~Brandenburg,\altaffilmark{3,4}
S.~C.~Cowley,\altaffilmark{4,5,6}
J.~L.~Maron,\altaffilmark{7}\\
and J.~C.~McWilliams\altaffilmark{8}}
\altaffiltext{1}{DAMTP, University of Cambridge, 
Wilberforce Road, Cambridge~CB3~0WA, UK; 
as629@damtp.cam.ac.uk.}
\altaffiltext{2}{Department of Physics, Norwegian University of Science 
and Technology, H{\o}yskoleringen 5, N-7034 Trondheim, Norway.}
\altaffiltext{3}{NORDITA, Blegdamsvej 17, DK-2100 Copenhagen \O, Denmark.}
\altaffiltext{4}{Isaac Newton Institute for Mathematical Sciences, 
20~Clarkson Road, Cambridge CB3 0EH, UK.} 
\altaffiltext{5}{Department of Physics and Astronomy, 
UCLA, Los Angeles, CA~90095-1547.} 
\altaffiltext{6}{Plasma Physics Group, Imperial College London, 
Blackett Laboratory, Prince Consort Road, London SW7 2BW, UK.}
\altaffiltext{7}{Department of Astrophysics, American Museum of Natural History, 
West 79th Street, New York, NY 10024-5192.}
\altaffiltext{8}{Department of Atmospheric Sciences, 
UCLA, Los Angeles, CA 90095-1565.}

\begin{abstract}

We study numerically the dependence of the critical magnetic 
Reynolds number~$\Rmc$ for the turbulent small-scale dynamo 
on the hydrodynamic Reynolds number~$\Re$. 
The turbulence is statistically homogeneous, isotropic, and mirror--symmetric.
We are interested in the regime of low magnetic Prandtl number $\Pm=\Rm/\Re<1$, 
which is relevant for stellar convective zones, protostellar 
disks, and laboratory liquid-metal experiments. 
The two asymptotic possibilities are $\Rmc\to\const$ as $\Re\to\infty$ 
(a small-scale dynamo exists at low $\Pm$) 
or $\Rmc/\Re=\Pmc\to\const$ as $\Re\to\infty$ 
(no small-scale dynamo exists at low $\Pm$).
Results obtained in two independent sets of simulations of 
MHD turbulence using grid and spectral codes are brought together 
and found to be in quantitative agreement. 
We find that at currently accessible resolutions, 
$\Rmc$ grows with $\Re$ with no sign of approaching a constant limit. 
We reach the maximum values of $\Rmc\sim500$ for $\Re\sim3000$. 
By comparing simulations with Laplacian viscosity, 
fourth-, sixth-, and eighth-order hyperviscosity and Smagorinsky large-eddy 
viscosity, we find that $\Rmc$ is not sensitive to the particular 
form of the viscous cutoff. 
This work represents a significant extension of the studies 
previously published by 
Schekochihin et al. (2004a) 
and Haugen et al. (2004a) 
and the first detailed scan of the 
numerically accessible part of the stability curve $\Rmc(\Re)$. 
\end{abstract}


\section*{}
\vskip-0.55cm

The magnetic Prandtl number~$\Pm$, which is the ratio of the kinematic 
viscosity to the magnetic diffusivity, is a key parameter of 
MHD turbulence. In fully ionized plasmas, 
$\Pm \approx 2.6\times10^{-5} T^4/n$,
where $T$ is the temperature in kelvins and $n$ the ion number density 
in units of cm$^{-3}$. In hot rarefied plasmas, such as the warm and hot phases 
of the interstellar medium or the intracluster medium, 
$\Pm\gg1$. In contrast, in the Sun's convective zone, 
$\Pm\sim10^{-7}$ to $10^{-4}$, in planets, $\Pm\sim10^{-5}$, 
and in protostellar disks, while estimates vary, 
it is also believed that $\Pm\ll1$ 
\citep[e.g.,][]{Brandenburg_Subramanian}. 
All these astrophysical bodies have disordered 
fluctuating small-scale magnetic fields and, in some cases, 
also large-scale ``mean'' fields. As they also have large 
Reynolds numbers and large-scale sources of energy, 
they are expected to be in a turbulent state. 
It is then natural to ask if their magnetic fields 
are a product of turbulent dynamo. 

To be precise, there are two types of dynamo. The large-scale, 
or {\em mean-field dynamo} generates magnetic fields at scales larger 
than the energy-containing scale of the turbulence, as is, for example, the case 
in helical turbulence. {\em The small-scale dynamo} amplifies magnetic 
fluctuation energy below the energy-containing scale of the turbulence. 
The small-scale dynamo is due to random stretching of the 
magnetic field by turbulent motions and does not depend on the 
presence of helicity. Mean-field dynamos 
typically predict field growth on time scales associated with the 
energy-containing scale (or longer), 
while the small-scale dynamo 
amplifies magnetic energy at the turbulent rate of stretching. 
Thus, the small-scale dynamo is usually a much faster process than 
the mean-field dynamo, and the large-scale field produced 
by the latter can be treated as approximately constant 
on the time scale of the small-scale dynamo.
The mean-field dynamo (or, more generally, a large-scale 
magnetic field of any origin) also gives rise to 
small-scale magnetic fluctuations because of the turbulent shredding 
of the mean field: this leads to algebraic-in-time growth 
of the small-scale magnetic energy --- again, 
a slower generation process than the exponential-in-time 
small-scale dynamo. 

In the systems with $\Pm\gg1$, the existence of the small-scale 
dynamo is well established numerically and has a solid theoretical 
basis (see \citealt{SCTMM_stokes} for an account of the relevant 
theoretical and numerical results and for a long list of 
references). The situation is much less well understood for 
the case of small~$\Pm$. It has been largely 
assumed that a small-scale dynamo should be operative in this regime 
as well. For example, the presence of large amounts of small-scale 
magnetic flux in the solar photosphere \citep[e.g.,][]{Title_review} 
has been attributed to small-scale dynamo action. This appeared 
to be confirmed by numerical simulations of the MHD turbulence 
in the convective zone \citep{Cattaneo_solar,Cattaneo_Emonet_Weiss,Nordlund_book}.
However, such simulations are usually done at $\Pm\ge1$ 
($\Pm=5$ in Cattaneo's simulations). 
Previous attempts to simulate MHD turbulence in various contexts 
with $\Pm<1$ found achieving dynamo in this regime much more difficult 
than for $\Pm\ge1$ \citep{Nordlund_etal,Brandenburg_etal_structures,Nore_etal,Christensen_Olson_Glatzmaier,MCM}. 
A systematic numerical investigation of the effect of $\Pm$ 
on the efficiency of the small-scale dynamo was 
carried out by \citet{SCMM_lowPr}, who found that 
the critical magnetic Reynolds number $\Rmc$ required for the small-scale 
dynamo to work increases sharply at $\Pm<1$. 
An independent numerical study by \citet{HBD_pre} confirmed this result. 

What are the basic physical considerations that should guide us 
in interpreting this result? 
First of all, let us stress that all working numerical 
small-scale dynamos are of the large-$\Pm$ kind 
(the case of $\Pm=1$ is nonasymptotic, but its properties 
that emerge in numerical simulations suggest that it belongs to the same class). 
Two essential features of the large-$\Pm$ dynamos are (1) the scale 
of the velocity field is larger than the scale of the magnetic 
field, and (2) the velocity field that drives the dynamo is 
spatially smooth and locally looks like a random 
linear shear, so the dynamo is due to exponential-in-time 
separation of Lagrangian trajectories and the consequent 
exponential stretching of the magnetic field. 
The basic physical picture of such 
dynamos (\citealt{Zeldovich_etal_linear}; see 
discussion in \citealt{SCTMM_stokes}; see also 
a review of an alternative but complementary approach 
by \citealt{Ott_review}) 
explicitly requires these two conditions to hold. 
The map dynamos and the dynamos in deterministic chaotic 
flows that were extensively studied in the 1980s--1990s 
\citep[see review by][]{STF} are all of this kind. 
The numerical dynamos with $\Pm\ge1$ 
(the first due to \citealt{Meneguzzi_Frisch_Pouquet})
are of this kind 
as well because they are driven by the spatially smooth 
viscous-scale turbulent eddies, which have the largest turnover rate. 

\pseudofigureone{fig_sketch}{Rmc_sketch.epsf}{fig1.epsf}{Sketch of 
the two possible shapes of the stability curve $\Rmc$ vs.\ $\Re$ for 
the small-scale dynamo.}

When $\Pm\ll1$ with both $\Rm\gg1$ and $\Re\gg1$, 
the characteristic scale $\lB$ of the magnetic field lies 
in the inertial range. For Kolmogorov turbulence, a simple estimate gives 
$\lB\sim\Rm^{-3/4}\lf$, where $\lf$ is the energy-containing scale. 
As the viscous scale is $\ld\sim\Re^{-3/4}\lf$, 
we have $\lf\ll\lB\ll\ld$. 
In a rough way, one can think of the turbulent eddies 
at scales $l>\lB$ as stretching the field at the 
rate $u_l/l$ and of the eddies at scales 
$l<\lB$ as diffusing the field with the 
turbulent diffusivity $u_l l$. In Kolmogorov 
turbulence, $u_l\sim l^{1/3}$, so both the dominant 
stretching and the dominant diffusion are due to the 
eddies at scale $l\sim\lB$. The resulting rates of 
stretching and of turbulent diffusion are of the same 
order, so the outcome of their competition cannot be determined 
on this qualitative level \citep{Kraichnan_Nagarajan}. 
An important conclusion, however, can be drawn. 
If the bulk of the magnetic energy is at the scale $\lB$,
the existence of the dynamo is entirely decided 
by the action of the velocities at the scale~$\lB$. 
Then it cannot matter 
where in the inertial range $\lB$ lies. But $\lB/\ld\sim\Pm^{-3/4}$, 
so the value of $\Pm$ does not matter as long as it is asymptotically 
small. Therefore, there are two possibilities: either 
there is a dynamo at low $\Pm$ and $\Rmc\to\const$ as $\Re\to\infty$ 
or there is not and there exists a finite 
$\Pmc=\Rmc/\Re\to\const$ as $\Re\to\infty$. 
Strictly speaking, the third possibility is that $\Rmc\propto\Re^\alpha$, 
where $\alpha$ is some fractional power, but this can only happen 
if the intermittency of the velocity field (non--self-similarity 
of the inertial range) is important for the existence of the 
dynamo.\footnote{The role of coherent structures can be 
prominent in quasi--two-dimensional dynamos 
(three-component velocity field depending on two 
spatial variables), where the inverse cascade characteristic of 
the two-dimensional turbulence gives rise to persistent 
large-scale vortices, which drive the dynamo 
\citep{Smith_Tobias}.}

The two possibilities identified above are illustrated in 
\figref{fig_sketch}. Several authors 
\citep{Vainshtein_lowPr,Rogachevskii_Kleeorin,Boldyrev_Cattaneo} 
showed that, given certain reasonable assumptions, 
the first possibility ($\Rmc\to\const$) 
is favored by the \citet{Kazantsev} model: the small-scale dynamo 
in a Gaussian white-in-time velocity field. 
In particular, \citet{Boldyrev_Cattaneo} found that 
the Kazantsev model gives $\Rmc$ 
that is roughly 10 times larger in the $\Pm\ll1$ 
regime than in the $\Pm\gg1$ regime 
(\citealt{Rogachevskii_Kleeorin} predict $\Rmc\sim400$, 
which is consistent with that). 
This prompted them to declare the issue settled on the grounds that 
the failure of the dynamo in numerical experiments at current 
limited resolutions is compatible with such an increase in $\Rmc$. 
However, the $\Pm\ll1$ dynamo in the Kazantsev model is 
a quantitative mathematical consequence of the model, 
and it is not known if and how it is affected 
by such drastic and certainly unrealistic assumptions as 
the Gaussian white-noise statistics for the velocity 
field.\footnote{\citet{Vainshtein_Kichatinov} 
argue that the equations that arise from the Kazantsev model are valid 
for non-white velocity fields if 
$n$-point joint probability density functions of Lagrangian 
displacements satisfy Fokker-Planck equations with some diffusion 
tensor. They further assume (on dimensional grounds) 
that this diffusion tensor scales as the scale-dependent turbulent diffusion 
$\sim u_l l$. This is, in fact, a closure scheme that we believe to be 
equivalent to Kazantsev's zero-correlation-time theory.} 
The existence of a dynamo in real turbulence is also 
a quantitative question (see discussion above), so it 
cannot be decided by a model that is not a quantitative 
approximation of turbulence. 

Thus, the issue cannot be considered settled until definitive 
numerical evidence is produced. This is an especially hard task 
because we do not know just how high a magnetic 
Reynolds number we must achieve in order to clearly 
see the distinction between $\Rmc\to\const$ and $\Rmc/\Re\to\const$. 
In this Letter, we have collected numerical results 
from two independent computational efforts: simulations 
using an incompressible spectral MHD code 
\citep[see code description in][]{Maron_Goldreich,MCM} 
and weakly compressible simulations using a grid-based  
high-order MHD code (the {\sc Pencil Code}\footnote{See 
http://www.nordita.dk/software/pencil-code.}). 

The equations we solved numerically 
(in a triply periodic cube) are 
\bea
\label{NSEq}
\dd_t\vu + \vu\cdot\vdel\vu &=& -{\vdel p\over\rho} 
+ {(\vdel\times\vB)\times\vB\over4\pi\rho} + \vF + \vf,\\
\dd_t\vB &=& \vdel\times\lt(\vu\times\vB\rt) + \eta\nabla^2\vB,
\label{eq_B}
\eea
where $\vu$ is the velocity and $\vB$ is the magnetic field 
(the {\sc Pencil Code}, in fact, solves the evolution equation 
for the vector potential ${\bf A}$ 
and then computes $\vB=\vdel\times{\bf A}$). 
All runs reported below are in the kinematic regime, 
$|\vB|\ll|\vu|$, so the Lorentz force in \eqref{NSEq} plays no role. 
Turbulence is driven by a random white-in-time nonhelical  
body force $\vf$ concentrated at $k=\kf$, where $\kf$
is the wavenumber associated with the box size. 
The (hyper)viscous force is 
\bea
\vF = {1\over\rho}\vdel\cdot\bl[2\<\rho\>\nu_n\lt(-\nabla^2\rt)^{n-1}\vS\br],
\label{def_F}
\eea
where $\nu_n$ is the fluid viscosity and 
\bea
S_{ij} = {1\over2}\lt({\dd u_i\over\dd x_j} + {\dd u_j\over\dd x_i}\rt) 
- {1\over3}\,\delta_{ij}\vdel\cdot\vu.
\eea
In the spectral simulations, 
the density $\rho=1$, and the incompressibility constraint 
$\vdel\cdot\vu=0$ is enforced exactly via the  
determination of the pressure~$p$. 
The grid simulations are isothermal: $p=\cs^2\rho$
with sound speed $\cs=1$, and the density satisfies 
\bea
\dd_t\rho + \vdel\cdot\lt(\rho\vu\rt) = 0.
\eea
We stay in the weakly compressible regime of low Mach numbers 
$M=\usq^{1/2}/\cs\sim 10^{-1}$ and $\rho\simeq\<\rho\>=1$
(angular brackets denote volume averages). 
Some numerical results on the onset of dynamo action at 
larger Mach numbers are given in \citet{Haugen_Brandenburg_Mee}.

The dissipation in the induction equation~\exref{eq_B} is 
always Laplacian with magnetic diffusivity~$\eta$ 
(we choose not to tamper with magnetic dissipation because 
we are interested in the sensitive question of field 
growth or decay). 
With regard to the viscous dissipation, we perform three 
kinds of simulations:

1.~Simulations with Laplacian viscosity: $n=1$ in \eqref{def_F}.

2.~Simulations with fourth-, sixth-, and eighth-order hyperviscosities: 
$n=2$, $3$, and~$4$, respectively, in \eqref{def_F}.

3.~Large-eddy simulations (LES) with the Smagorinsky 
effective viscosity \citep[e.g.,][]{Pope_book}: 
in \eqref{def_F}, $n=1$, and $\nu_1$ is replaced by
$\nu_{\rm S} = \bl(C_{\rm S}\Delta\br)^2\bl(2\vS:\vS\br)^{1/2}$,
where $\Delta$ is the mesh size 
and $C_{\rm S}=0.2$ is an empirical coefficient. 

The magnetic Reynolds number is defined by
$\Rm=\usq^{1/2}/\kf\eta$, where $\kf$ is the box wavenumber 
(the smallest wavenumber in the problem). 
For the runs with Laplacian viscosity ($n=1$), 
the hydrodynamic Reynolds number is 
$\Re=\usq^{1/2}/\kf\nu_1$. 
For hyperviscous runs and for LES, we define $\Re$ 
by replacing $\nu_1$ with the effective viscosity: 
\bea
\label{nueff_def}
\nueff=\epsilon/\<2\vS:\vS\> = \epsilon/\dusq 
\eea
(the second expression is for the spectral simulations, 
where $\vdel\cdot\vu=0$ exactly). 
Here $\epsilon=\<\vf\cdot\vu\>$ is the total injected power 
and is equal to the total energy dissipation. 
As the forcing $\vf$ is white in time, $\epsilon=\const$: 
indeed, given 
$\<f^i(t,\vx)f^j(t',\vx')\> = \delta(t-t')\,\epsilon^{ij}(\vx-\vx')$, 
it is easy to show that $\epsilon=\frac{1}{2}\,\epsilon^{ii}(0)$. 

The results of all our simulations are presented in \figref{fig_Rmc}, 
where $\Rmc$ is plotted versus $\Re$. 
Each value of $\Rmc$ was computed by interpolating 
between least-squares--fitted growth/decay rates of a growing and a decaying run. 
Error bars are based on $\Rm$ and $\Re$ for these pairs of runs. 
The only exception is the point enclosed in a circle, 
which corresponds to $(\Rm,\Re)$ for a run that appeared 
to be marginal (in this case we could not afford the resolution necessary 
to achieve a decaying case). The run times in all cases 
were long enough for the least-squares--fitted growth rates to stop 
changing appreciably (typically this required about $20$ box-crossing 
times, but cases close to marginal needed longer running times). 

\pseudofigureone{fig_Rmc}{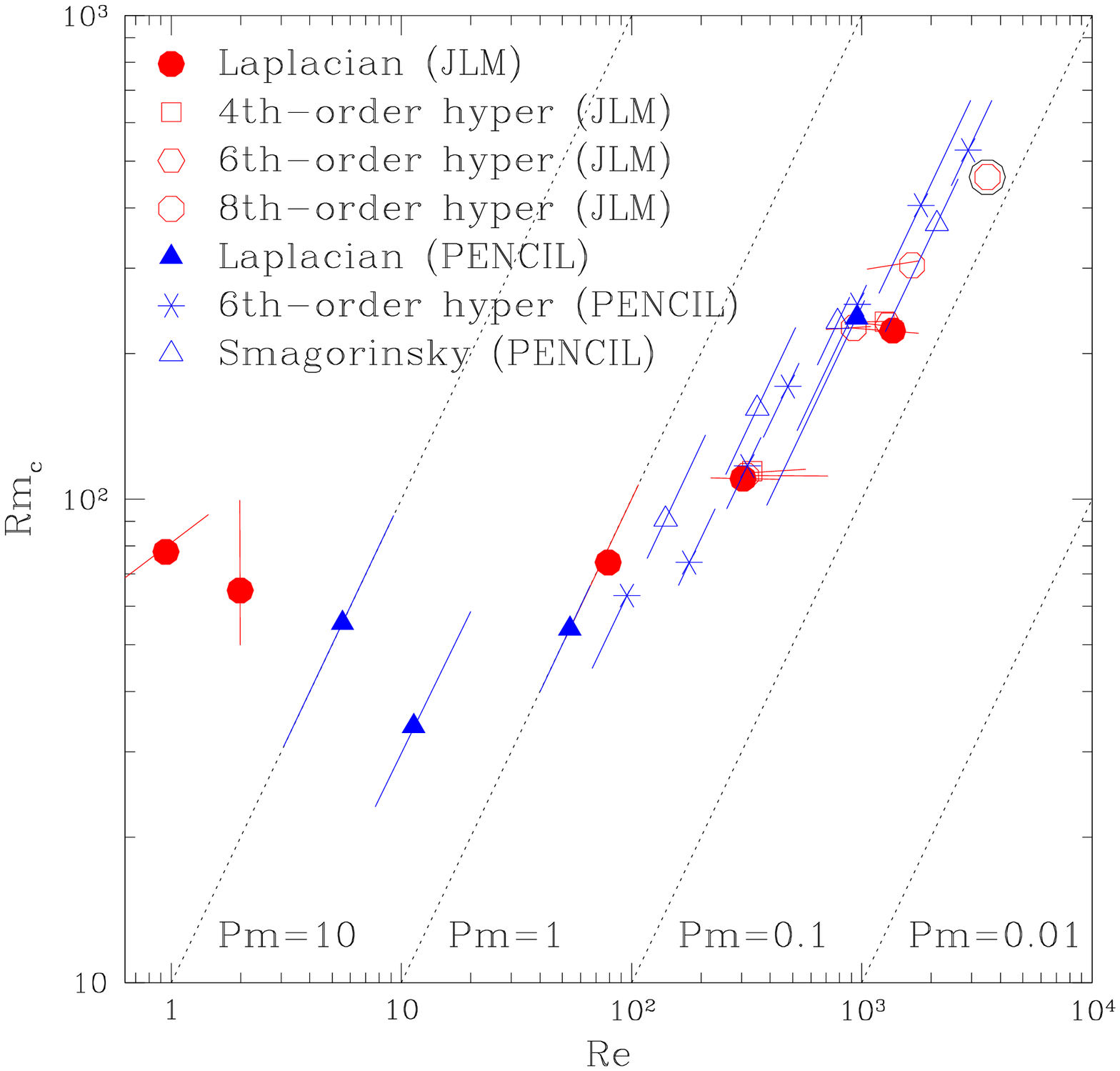}{fig2.ps}{Dependence of $\Rmc$ on $\Re$. 
``JLM'' refers to simulations done with the incompressible spectral code 
written by J.~L.~Maron: runs with Laplacian viscosity, fourth-, sixth-, 
and eighth-order 
hyperviscosity (resolutions $64^3$--$256^3$). In this set 
of simulations, hyperviscous runs were done at the same values of $\eta$ 
as the Laplacian runs, so the difference between the results for these runs 
is nearly imperceptible. 
``PENCIL'' refers to weakly compressible simulations done with the {\sc Pencil Code}: 
runs with Laplacian viscosity, sixth-order hyperviscosity, and 
Smagorinsky large-eddy viscosity (resolutions $64^3$--$512^3$).}

We see that there is good agreement between 
the results for runs with different forms of viscous dissipation; 
this confirms the natural assumption
that the field-generation properties 
of the turbulence at low $\Pm$ are not sensitive to the 
way the velocity spectrum is cut off. 
It is also encouraging that results from 
two very different codes are in quantitative agreement. 

Our previous studies \citep{SCMM_lowPr,HBD_pre} 
had the maximum value of $\Rmc\sim200$. The results 
reported here raise it to $\sim500$, with the corresponding 
values of $\Pmc$ around 0.15. While a roughly 10-fold increase with 
respect to $\Rmc$ for the $\Pm=1$ dynamo has now been achieved, 
there is thus far no sign of $\Rmc$ reaching an asymptotically 
constant value. 
This said, the current resolutions 
are clearly still insufficient to make a definitive judgement, 
although we are now very close to values of $\Rmc$ predicted by the 
theories based on the Kazantsev model 
\citep{Rogachevskii_Kleeorin,Boldyrev_Cattaneo} 
--- whether or not the model yields quantitatively correct predictions 
should become clear in the near future. 

The numerical results reported above concerned the 
dependence $\Rmc(\Re)$ for 
the turbulent small-scale dynamo, i.e., the ability of 
turbulent velocity fluctuations to amplify
magnetic energy at scales smaller than the 
energy-containing scale of the turbulence. 
The $\Rmc(\Re)$ dependence is also an interesting 
issue for other kinds of dynamo. 

If the velocity field is non--mirror-symmetric, it can often 
drive the mean-field dynamo (MFD), which means the growth 
of the magnetic field at scales larger than the energy-containing scale 
of the turbulence \citep{Krause_Raedler}. 
This large-scale field generated by the MFD, 
just like a mean field imposed externally, 
can induce small-scale magnetic fluctuations 
as it is shredded by the turbulence, so the total 
field has both a mean (large-scale) and a fluctuating component. 
In many cases, the breaking 
of the mirror symmetry leads to a non-zero value of the 
net helicity $\<\vu\cdot(\vdel\times\vu)\>\neq0$
(the average is over all scales smaller 
than the energy-containing scale of the turbulence, 
{\em not} over the entire volume of the system). 
The mean-field generation is then referred to as 
the $\alpha$-effect. The stability curve $\Rmc(\Re)$ for the 
$\alpha$-effect is different than for the small-scale dynamo: 
it is essentially a condition for at least one unstable 
large-scale mode to fit into the system. In a numerical study done 
with the same code as the grid simulations reported above 
but with fully helical random forcing, \citet{Brandenburg_alpha} 
found much lower values of $\Rmc$ than for the small-scale 
dynamo and very little dependence of $\Rmc$ on $\Pm$ 
for $\Pm\ge0.1$. 

A non-zero net helicity is not 
a necessary condition for the MFD \citep[e.g.,][]{Gilbert_Frisch_Pouquet}. 
In fact, it has been suggested recently by \citet{Rogachevskii_Kleeorin_sc}
that the MFD can be driven simply by the presence of a constant 
mean velocity shear (shear-current or $\delta$-effect) 
--- a very generic possibility of obvious relevance 
to systems with mean flows. Mean flows are present 
in many astrophysical cases and in all current laboratory dynamo experiments 
\citep{Gailitis_etal,Mueller_Stieglitz_Horanyi,Bourgoin_etal,Lathrop_Shew_Sisan,Forest_etal}. 
A mean flow can be a dynamo 
in its own right: an MFD (field growth at scales above the flow scale)
and, if the flow has chaotic trajectories 
in three dimensions, also a small-scale dynamo 
(field growth at scales $\sim\Rm^{-1/2}$ times the scale of the flow; 
see \citealt{STF} --- as noted above, small-scale dynamos in 
deterministic chaotic flows are equivalent to the large-$\Pm$ case). 
When $\Re$ is large, the energy of the turbulent velocity fluctuations
is comparable to the energy of the mean flow. 
The critical $\Rm$ required for field growth will have some 
dependence on $\Re$, which reflects the effect of the turbulence 
on the structure of the mean flow and/or on the effective value 
of the magnetic diffusivity \citep[the $\beta$-effect; see][]{Krause_Raedler}. 
This dependence was the subject of two recent numerical studies: 
of the dynamo in a turbulence with 
a constant Taylor-Green forcing by \citet{Ponty_etal}, 
and of the Madison dynamo experiment 
(propeller driving in a spherical domain) by \citet{Bayliss_Forest}. 
The $\Re$ dependence of $\Rmc$ that emerges from such 
simulations is distinct from that for a pure small-scale dynamo.
Indeed, Y.~Ponty et al.~(2005, private communication) have shown that, 
in the limit of large $\Re$, 
the value of $\Rmc$ in their simulations tends to a constant 
that coincides with $\Rmc$ 
calculated for the mean flow alone, i.e., for the velocity field 
with fluctuations removed by time averaging. 
In contrast, the subject of the present Letter has been 
the possibility of a small-scale dynamo driven solely by turbulent 
fluctuations, in the absence of a mean flow. The importance 
of this possibility or lack thereof is that such a dynamo, 
if it exists, occurs at the turbulent stretching rate 
associated with the resistive scale. 
This is much faster (by a factor of $\sim\Rm^{1/2}$; see discussion above)
than the growth rate of any MFD or of a small-scale dynamo associated with 
the mean flow, i.e., than the stretching 
rate at the energy-containing scale or at the scale of the mean flow. 

\acknowledgements

It is a pleasure to acknowledge discussions with 
participants of the program ``Magnetohydrodynamics of Stellar Interiors'' 
at the Isaac Newton Institute, Cambridge (UK), 
especially S.~Fauve, N.~Kleeorin, Y.~Ponty, M.~Proctor, and I.~Rogachevskii. 
We also thank C.~Forest and J.-F.~Pinton for discussions 
of both laboratory and numerical dynamos. Simulations were done 
at the UKAFF (Leicester), NCSA (Illinois), 
Norwegian High Performance Computing Consortium (Trondheim and Bergen), 
and the Danish Center for Scientific Computing. 
This work was supported in part by NSF grant AST~00-98670 
and by the US DOE Center for Mutiscale Plasma Dynamics. 
A.A.S.\ was supported by 
the UKAFF Fellowship. 
N.E.L.H.\ was supported in part by the David Crighton 
Visiting Fellowship (DAMTP, Cambridge).

\end{document}